\documentclass[fleqn,usenatbib]{mnras}

\usepackage{newtxtext,newtxmath}

\usepackage[T1]{fontenc}

\DeclareRobustCommand{\VAN}[3]{#2}
\let\VANthebibliography\thebibliography
\def\thebibliography{\DeclareRobustCommand{\VAN}[3]{##3}\VANthebibliography}

\usepackage{graphicx}	
\usepackage{amsmath}	
\usepackage{url}
\usepackage[none]{hyphenat}

\title[Autoencoder Spectral Fitting]{Rapid Spectral Parameter Prediction for Black Hole X-Ray Binaries using Physicalised Autoencoders}

\author[E. Tregidga et al.]{
Ethan Tregidga,$^{1,2}$\thanks{E-mail: et1g19@soton.ac.uk}
James F. Steiner,$^{1}$
Cecilia Garraffo,$^{1}$
Carter Rhea,$^{3,4}$
Mayeul Aubin$^{1,5}$
\\
$^{1}$Center for Astrophysics | Harvard \& Smithsonian, 60 Garden Street, Cambridge, MA 02138, USA\\
$^{2}$University of Southampton, University Road, Southampton, SO17 1BJ, UK\\
$^{3}$D\'{e}partement de Physique, Universit\'{e} de Montr\'{e}al, Succ. Centre-Ville, Montr\'{e}al, Qu\'{e}bec, H3C 3J7, Canada\\
$^{4}$ Centre de recherche en astrophysique du Québec (CRAQ), Qu\'{e}bec QC G1V 0A6, Canada\\
$^{5}$\'Ecole Polytechnique, Rte de Saclay, 91120 Palaiseau, France
}

\pubyear{2024}

\begin{document}
\label{firstpage}
\pagerange{\pageref{firstpage}--\pageref{lastpage}}
\maketitle

\begin{abstract}
Black hole X-ray binaries (BHBs) offer insights into extreme gravitational environments and the testing of general relativity.
The X-ray spectrum collected by NICER offers valuable information on the properties and behaviour of BHBs through spectral fitting.
However, traditional spectral fitting methods are slow and scale poorly with model complexity.
This paper presents a new semi-supervised autoencoder neural network for parameter prediction and spectral reconstruction of BHBs, showing an improvement of up to a factor of 2,700 in speed while maintaining comparable accuracy.
The approach maps the spectral features from the numerous outbursts catalogued by NICER and generalises them to new systems for efficient and accurate spectral fitting.
The effectiveness of this approach is demonstrated in the spectral fitting of BHBs and holds promise for use in other areas of astronomy and physics for categorising large datasets.
The code is available via \url{https://github.com/EthanTreg/Spectrum-Machine-Learning}.
\end{abstract}

\begin{keywords}
accretion, accretion discs -- black hole physics -- methods: data analysis -- X-rays: binaries
\end{keywords}



\section{Introduction}

Black hole X-ray binaries (BHBs) consist of a black hole (BH) that accretes matter from a nearby companion star, releasing substantial amounts of energy as X-ray photons through gravitational potential energy loss \citep{novikov1973astrophysics}. 

Two properties, mass (M) and dimensionless spin ($a_*$), uniquely define a BH in general relativity via the Kerr metric.
The mass of the BH can be determined empirically using the mass function of a binary system and the radial velocity \citep{casares2014mass}.
The value of $a_*$ can range in magnitude from zero, representing a non-spinning Schwarzschild hole, to one, representing a maximally rotating Kerr hole \citep{remillard2006x}, with negative values representing a counter-rotating BH relative to the accretion disk \citep{shapiro1976black}.
However, there is a practical maximum limit of $|a_*|\simeq 0.998$ caused by a counteracting torque from the photons captured by the BH as photons with negative angular momentum have a larger cross-section than photons with positive angular momentum \citep{thorne1974disk}.

BHs have two characteristic radii: the event horizon, where the radial escape velocity equals the speed of light, making all information inside inaccessible to outside observers, and the innermost stable circular orbit (ISCO), which is the smallest stable orbit for orbiting matter, usually in the form of an accretion disk \citep{remillard2006x, heinicke2015schwarzschild}.
As $a_*$ becomes larger and more positive, the ISCO and event horizon shrink and the ISCO radius approaches the event horizon radius.

The gas accreted from the companion star circularises and, via viscous drag forces, dissipates energy while exchanging angular momentum, forming an accretion disk which, during soft states, section~\ref{sec:spectra}, is observed to extend to the ISCO, or very proximate to it \citep{cui1997evidence, davis2006testing}.

X-ray spectral analysis provides important constraints on the properties and evolving characteristics of the accretion flow as well as the BH's properties  (e.g., mass accretion rate $\dot{M}$ and $a*$, e.g., \citealt{fabian2014determination, zhao2021re}).



\subsection{Black Hole X-Ray Binary Spectra}
\label{sec:spectra}

\begin{figure}
    \centering
    \includegraphics[width=\columnwidth]{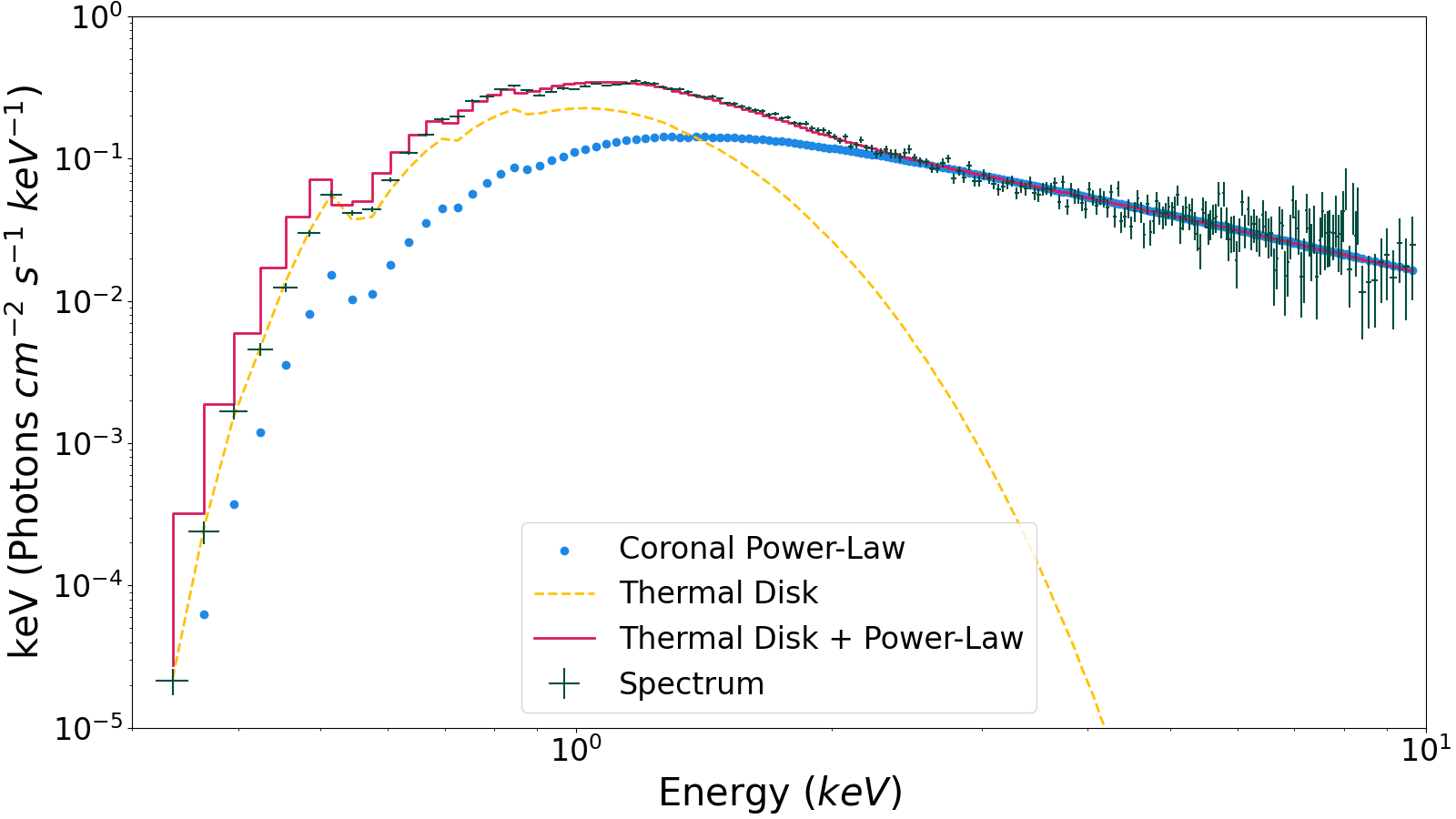}
    \caption{Energy unfolded spectrum of a BHB (green plus) fitted with the model (red line): {\em TBabs(simplcut$\otimes$ezdiskbb)} showing the thermal disk (yellow dashed) and coronal power-law components (blue dots).}
    \label{fig:spectrum}
\end{figure}

As seen during outburst episodes in which the inflow of matter onto the BH evolves over time, BHBs show three primary X-ray spectral continuum components.
Figure~\ref{fig:spectrum} shows an energy unfolded spectrum (green plus) fitted with the model {\em TBabs(simplcut$\otimes$ezdiskbb)} (red line) from the {\sc PyXspec} software \citep{arnaud1996xspec, gordon2021pyxspec}, see section~\ref{sec:method}.
(1) The thermal component ({\em ezdiskbb}, yellow dashed) is modelled by a $\sim 1$~keV-peaked multi-temperature disk model (MTD) produced by a geometrically thin, optically thick accretion disk.
(2) The nonthermal component ({\em simplcut}, blue dots) is a power law produced by thermal photons Compton upscattering off hot electrons ($kT_e \sim 100$ keV) in an optically thin, or marginally-thick corona.
(3) The power-law component (blue dotted) extends to high energies and may exhibit a break or an exponential cutoff.
(4) The spectrum may also contain an additional nonthermal component referred to as reflection (not shown in Figure~\ref{fig:spectrum}), which is produced when Comptonised coronal photons illuminate the accretion disk and are reprocessed at its surface, generating a spectral bump between 20 to 30 keV \citep{esin1997advection, guilbert1988cold, remillard2006x, done2007modelling, gilfanov2009x, ingram2019review}.
The reflection of high-energy photons produces fluorescent excitation of elements in the accretion disk, with the Fe-K$\alpha$ emission line at $\sim 6.4$ keV being the most prominent.
Relativistic effects from the spin of the BH are imprinted on the skewed profile of the resulting Fe-K$\alpha$ emission line \citep{fabian1989x, bambi2013testing}.

BHBs can demonstrate significant variations throughout an outburst event (typically lasting many months from start to finish), and the prominence of the different components can change dramatically depending on the system's state \citep{remillard2006x}.
BHBs spend most of their time in quiescence at exceedingly low luminosities; however, during outbursts, they undergo rapid and dramatic changes in brightness and accretion state \citep{remillard2006x, dunn2010global}.
The accretion states follow several possible nomenclatures; broadly, five states are commonly identified.
There is a hard low luminosity state (LS) where the corona dominates the spectrum, a high soft state (HS) where the thermal blackbody is dominant, a very high state (VHS) which exhibits both thermal emission and a strong steep power law, intermediate states (INT) between LH and VHS/HS, and finally the aforementioned quiescent state \citep{esin1997advection, done2007modelling, remillard2006x}.

\subsection{Spectral Fitting}

The current spectral fitting approach involves convolving models with instrumental responses and brute-force curve fitting to achieve an acceptable fit with data \citep{arnaud1996xspec}.
However, analysis with this approach will become increasingly challenging as next-generation microcalorimeters \citep{tashiro2020status}, survey instruments \citep{kraft2022line}, and fast timing instruments \citep{ray2019strobe} produce high volumes of spectral data and unprecedented spectral resolution, requiring increased modelling complexity.

\cite{de2019spex} identify challenges and improvements of current methods.
The main challenge they identified was that the increased resolution reveals more spectral lines, requiring more complex atomic models that are difficult to accelerate without a loss in accuracy.
Another challenge is the storage and loading of response files, which can readily become limiting, as high-resolution spectra require significantly larger response matrices to encode the instrumental performance.
They implemented new file structures and binning methods to reduce file sizes, improve parallel performance, and reduce execution times; however, machine learning could remove the need for response files in spectral fitting outside of training in certain domains.

In a pathfinder effort, we investigate using an autoencoder architecture inspired by the success of neural networks in image and signal processing \citep{krizhevsky2017imagenet, oord2016wavenet, baldi2012autoencoders}.
Our approach aims to predict spectral parameters more efficiently.

Previous work by \cite{parker2022agn} is the most similar to our work, where they implemented a simple neural network to predict active galactic nuclei spectral parameters from simulated spectra of the X-ray Athena survey instrument \citep{meidinger2018wide}.
Their results show that neural networks can significantly reduce computational time while obtaining similar accuracies.
They show another advantage of employing neural networks is avoiding local minima that traditional spectral fitting can get stuck in.

\subsection{Machine Learning}

Machine learning is an implicit set of optimisation algorithms that minimise a score, known as a loss function, that measures the inaccuracy of the algorithm's outputs to the ground truth \citep{wang2020comprehensive}.
Three main learning methods are supervised, unsupervised and semi-supervised learning.

\textbf{Supervised learning} involves classifying or predicting values and requires prior knowledge of the ground truth labels for training.
While supervised learning can provide physical outputs, generating labels can be computationally expensive and require human input.

\textbf{Unsupervised learning} helps find patterns and group data or learn low-dimensional data representations.
It can take advantage of large quantities of unlabeled datasets; however, the outputs from the network can be hard to interpret.

\textbf{Semi-supervised learning} aims to improve supervised learning by extending the labelled dataset by inferring labels from unlabeled data or, for unsupervised learning, informing the network of pre-existing groups to improve the accuracy and physical interpretation of the network \citep{van2020survey, zhu2005semi}.

\subsection{Neural Networks}

Neural networks are a type of machine learning algorithm formed from differentiable functions with free weights, called neurons, where a group of neurons form a layer, which can have different types, such as linear, convolutional and recurrent.
The weights within a neural network are tuned by propagating the loss through the weights using differentiation to minimise the loss \citep{rumelhart1986learning}.

\textbf{Linear layers} are composed of neurons with output $y_i$ that calculate the weighted sum of the neurons in the previous layer, $x_i$, as shown in equation~\ref{eq:linear}, where $\theta_i$ is the bias of the $i$th neuron and $w_{ij}$ is the weight of $x_j$ for the $i$th neuron \citep{svozil1997introduction}.

\begin{equation}
    \label{eq:linear}
    y_i=\theta_i+\sum_j{w_{ij}x_j}
\end{equation}

\begin{figure}
    \centering
    \includegraphics[width=\columnwidth]{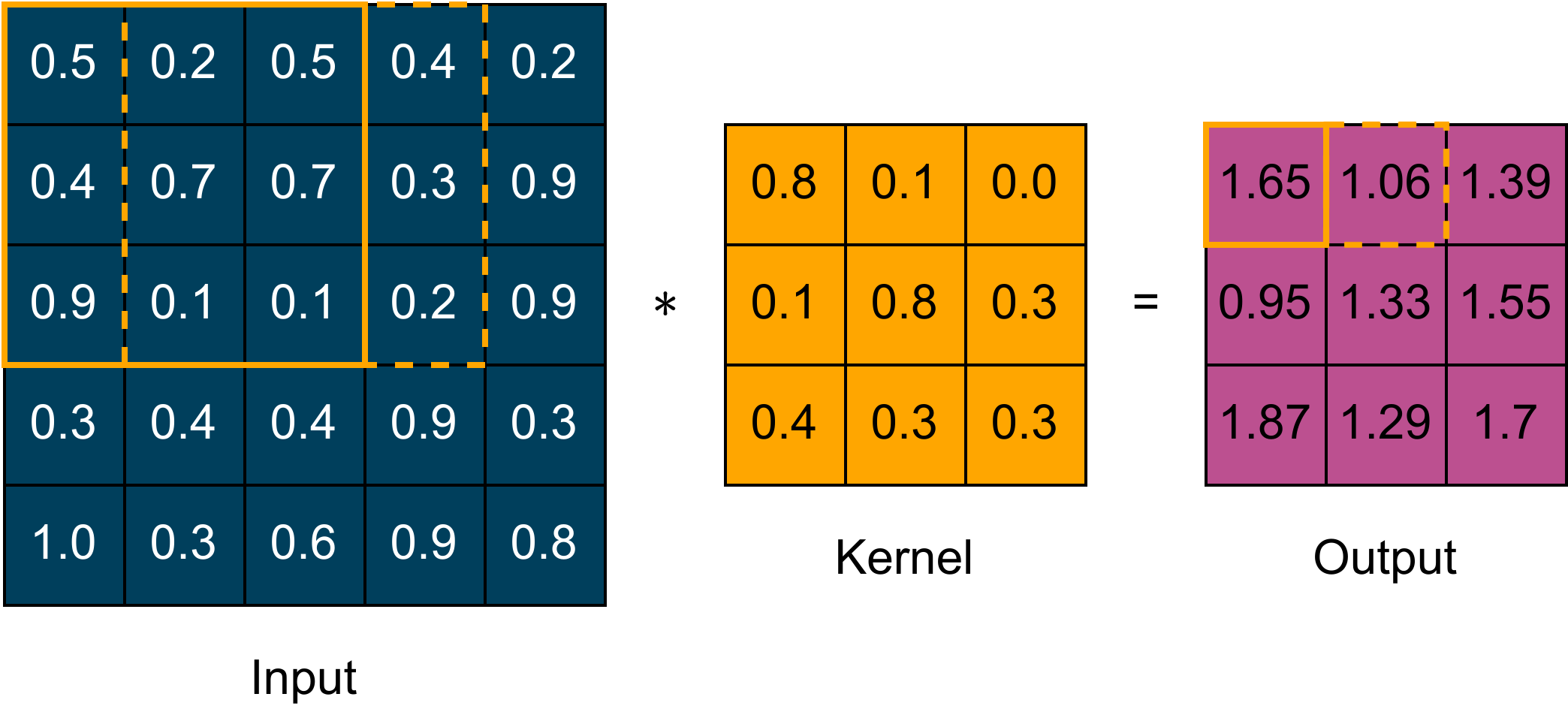}
    \caption{
        Convolutional layer with a $3\times3$ kernel of weights (orange kernel) convolved with the input (blue input) to produce an output (pink output) by calculating the weighted summation with the values within the receptive field of the kernel where the first step is shown by the orange box on the input, which produces the top left value in the output (1.65).
        The stride of the kernel indicates how many values the kernel moves between calculations, with the figure showing a stride of one.
        Therefore, the dashed orange box on the input shows the second step, which produces the top middle value (1.06).
    }
    \label{fig:convolution}
\end{figure}

\textbf{Convolutional layers} learn spatial features by sliding a kernel of weights across an input.
Figure~\ref{fig:convolution} shows the convolution of the input (blue input) with a $3\times3$ kernel of weights (orange kernel) to produce the output (pink output) by calculating the weighted sum of the inputs within the receptive field of the kernel at each step \citep{li2021survey}.
The stride defines the number of values the kernel moves between each step, so Figure~\ref{fig:convolution} shows a stride of one.
A convolutional layer can include several kernels, each optimising for a different feature set; therefore, the output can contain several channels of different feature maps.

\textbf{Recurrent layers} were introduced by \cite{hopfield1982neural} to learn sequential relationships from an input $\textbf{x}$, where the $t$th neuron in the layer produces an output, $y_t$, from the weighted sum of one value from the input, $x_t$ and the value from the output, $y_{t-1}$, of the $t-1$ neuron that used the previous value from the input sequence, $x_{t-1}$.
However, recurrent layers can suffer from vanishing gradients, in which the gradients go to zero, or conversely exploding gradients, which diverge to large values.
\cite{cho2014properties} proposed gated recurrent units, a modified recurrent network which improves gradient flow back to the input efficiently.

Stacking multiple layers allows neural networks to learn complex non-linear problems; however, this requires activation functions to introduce non-linearity between layers; otherwise, multilayer networks would collapse to a single layer \citep{rumelhart1986learning}.
The rectified linear unit activation function, $f(x)=\max(0,x)$, and its derivatives have become the most commonly used activation functions due to their success in deep neural networks \citep{glorot2011deep}; however, due to the zero gradient for negative values, rectified linear units can become dead, preventing the loss from propagating to the neurons closer to the input.
Exponential linear units modify the rectified linear unit to allow negative values, shown in equation~\ref{eq:elu} with $\alpha > 0$, preventing dead rectified linear units and improving training times, \citep{clevert2015fast}.
Scaled exponential linear units further modify the exponential linear unit by setting $\alpha \approx 1.6733$ and scaling the output by $\lambda \approx 1.0507$ to self-normalise the outputs to improve the stability of training \citep{klambauer2017self}.

\begin{equation}
    \label{eq:elu}
    \begin{cases}
    x & x \geq 0\\
    \alpha(e^x - 1) & x < 0
    \end{cases}
\end{equation}

\subsection{Autoencoders}

\begin{figure}
    \includegraphics[width=\columnwidth]{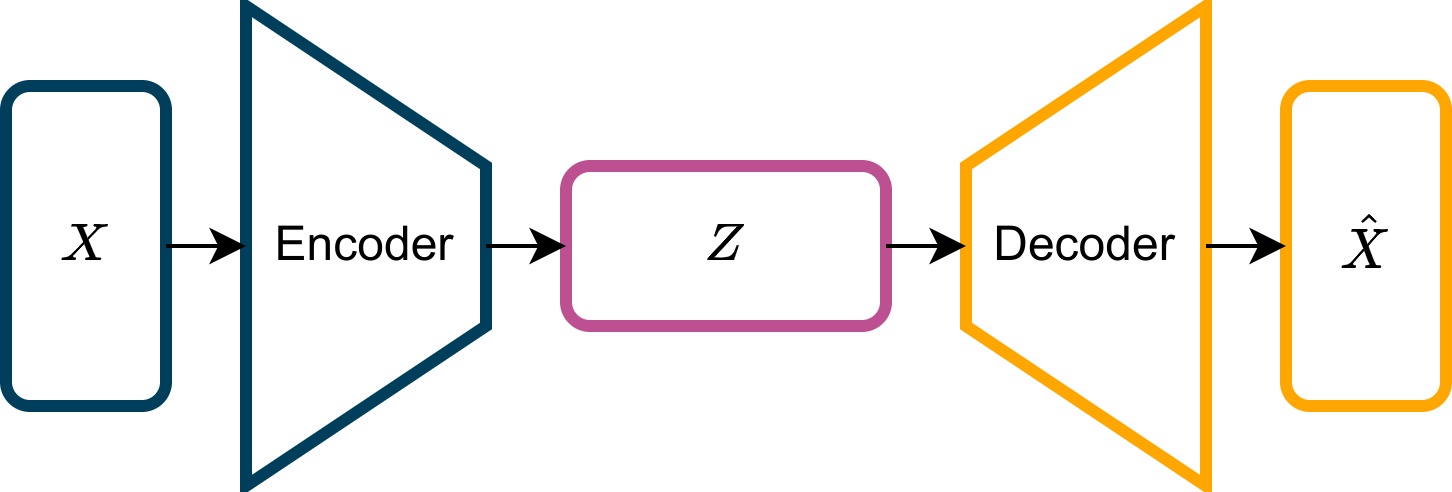}
    \caption{
        The general form of an autoencoder takes an input $x$, encodes it (blue) to a latent space $z$ (pink), and reconstructs the input $\hat{x}$ using a decoder (orange).
        The weights within the autoencoder are trained to minimise the difference between $\hat{x}$ and $x$.
        The autoencoder, therefore, learns to encode the most useful information required to reconstruct the input in the latent space.
    }
    \label{fig:basic_autoencoder}
\end{figure}

Autoencoders are a type of unsupervised method that compress high-dimensional data to a lower dimensional latent space (pink) using an encoder (blue) and then reconstruct the input from the latent space using a decoder (orange), as illustrated in Figure~\ref{fig:basic_autoencoder}.
The autoencoder is forced to encode the most important information into the latent space, and if the autoencoder is a single layer and the latent variables are orthogonal, then this is equivalent to principal component analysis \citep{ladjal2019pca}.
One drawback of autoencoders is that their latent space is generally non-physical; therefore, obtaining useful information from it can be challenging \citep{baldi2012autoencoders}.

Several techniques have been proposed for extracting useful information from the latent space.
For example, one approach is to use a portion of the latent space as variables that can be passed into a physical model, forcing those parameters to be physically grounded, as suggested by \cite{boone2021parsnip}.
Another technique is to train a classifier on the latent space using semi-supervised learning, as utilised by \cite{gogna2016semi, zemouri2020semi}.
Alternatively, \cite{ntampaka2022importance} designed their architecture to construct a physical latent space implicitly through physical reasoning of the latent structure and found that the latent variable correlated with the expected physical parameter.

\section{Method}
\label{sec:method}

In this study, we demonstrate a proof-of-concept neural network for predicting the parameters of BHB X-ray spectra, using a simple model as a starting point to test the feasibility of the network.
Specifically, we use the model {\em TBabs(simplcut$\otimes$ezdiskbb)} \citep{wilms2000absorption, steiner2017self, zimmerman2005multitemperature} in {\sc PyXspec} software.

The {\em ezdiskbb} model is an MTD radiation model that arises from the thermal emission of accretion disks, subject to the constraint of zero torque at the disk's inner edge \citep{zimmerman2005multitemperature}.
The model has two parameters: maximum accretion disk temperature ($kT_{\rm disk}$) and its normalisation ($N$), the latter of which is physically constraining of the disk inner-radius when several properties of the system, such as the disk inclination, source distance, and spectral hardness are accounted for (along with relativistic corrections).
We keep $kT_{\rm disk}$ and $N$ free for the network to learn.

{\em simplcut} is a convolution kernel that Compton upscatters a proportion of the MTD thermal photons in the hot corona.
Reprocessed coronal photons, which are reflected off the accretion disk, can be accounted for self consistently, conserving photons \citep{Steiner2009SIMPL}.
The model has four parameters: photon index ($\Gamma$); the fraction of thermal photons scattered in the corona ($f_{\rm sc}$); the proportion of Compton scattered photons that are reprocessed in the accretion disk ($R_F$) versus reaching the observer directly; and the temperature of electrons in the corona ($kT_e$).
We set $R_F = 0$\footnote{This is in effect equivalent to the simplifying approximation that any photons which return to the disk reflect elastically with no absorption or fluorescent reprocessing.} and $kT_e = 100$ keV as fixed parameters, allowing $\Gamma$ and $f_{\rm sc}$ to be free parameters.

The {\em TBabs} model describes X-ray absorption along the line of sight in the interstellar medium. It has a single free parameter, the hydrogen column density ($N_H$) \citep{wilms2000absorption}.

Our study aims to develop an encoder to generate spectral parameters from input spectra.
We enforce physical constraints on the latent space to overcome the limitations of an autoencoder's non-physical latent space by training the autoencoder in two steps.
The first is to train the decoder on labelled synthetic data using supervised learning, where the input is spectral parameters, and the output is reconstructed spectra.
Then, we train the encoder with unlabelled observations and the trained decoder, which conditions the latent space on the physical parameters.

Our model has two advantages by incorporating this physical constraint on the latent space.
Firstly, the encoder can learn to encode the physical parameters directly from the input spectra without first learning abstract representations.
Secondly, the decoder can rapidly create synthetic spectra given a set of physical parameters, which could be used to interpolate grids generated by computationally expensive models, such as XSTAR \citep{bautista2001xstar}, bypassing the model by emulating the output to accelerate fitting times.

\subsection{Data and Preprocessing}
\label{sec:preprocessing}

We obtained energy spectra for approximately 25 BHB systems from the Neutron Star Interior Composition Explorer (NICER).
NICER is a soft X-ray timing spectroscopic mission with its main instrument consisting of 52 active detectors, covering an energy range of $0.2 < E < 12$ keV and an energy resolution of $\sim 100$ eV \citep{gendreau2012neutron}.
NICER’s high temporal resolution and large collecting area enable the collection of useful spectra for BHBs (often $~10^4-10^7$ counts) in short observations ${\sim} 1000$ s, without being susceptible to pileup effects, allowing for flexible scheduling to collect large quantities of spectra.

We selected the entire catalogue of NICER BHB data as of August 2022 with an initial screening to remove any data overly contaminated by elevated background counts. We use each remaining continuous stare to produce a spectrum, yielding a total dataset of 10,800 spectra.
We used a custom pipeline using many components of the NICER data and analysis software suite\footnote{\url{https://heasarc.gsfc.nasa.gov/docs/software/lheasoft/help/nicer.html}} (NICERDAS) to process the raw data, which provided us with the counts against the channel for the spectrum and background and the corresponding redistribution matrix and auxiliary response matrix.
Additionally, we performed the following steps to standardise the data to ensure the neural network could learn the spectral features effectively:

\begin{enumerate}
    \item Exposure normalisation - Normalise the counts and background by the exposure time to obtain the detection rate per channel.
    \item Background correction - Correct the spectrum by subtracting the background rate. The 3C50 background model was used throughout \citep{Remillard2022}.
    \item Detector normalisation - Normalise the spectrum by the number of active detectors during the observation.
    \item Binning - Due to the channel energy resolution being higher than the detector energy resolution, we summed the data using the following bin sizes: 30 eV bins at 0.2 keV, 40 eV bins at 2.48 keV, 50 eV bins at 6 keV, and 60 eV bins up to 12 keV.
    Then, standardise the data to spectral density by dividing the spectrum by the width of each bin.
    \item Energy range - Limit the energy range to $0.3 < E < 10$ keV, over which NICER is most sensitive and corresponds to a typical spectral analysis range.
\end{enumerate}

We adopt a fixed binning to ensure a consistent data size, 240 bins for each spectrum.

However, the intensity of an individual spectrum and the differences between spectra can span many orders of magnitude, leading to challenges with network training, given the dynamic ranges at hand.
When using backpropagation to update the weights in the network, large inputs may result in the exploding gradient problem.
To address this, we transformed the spectra into logarithmic space and normalised all spectra between zero and one based on the amplitude range of synthetic spectra.
We also take the logarithm of the spectral parameters, except for photon index $\Gamma$, and normalise them to follow a normal distribution based on the synthetic parameters \citep{sola1997importance}.

\subsection{Synthesising Data}
\label{sec:synth_spectra}

We used the fakeit function in {\sc PyXspec} to create synthetic spectra for training based on the model {\em TBabs(simplcut$\otimes$ezdiskbb)}, with a fiducial exposure time of 1 ks.
We randomly select a spectrum from the real dataset to generate a set of 100 synthetic spectra.
We used its background, redistribution matrix, and auxiliary response files as the settings for the synthetic spectra.
We then randomly assigned values for each parameter in the model for each synthetic spectrum from either a logarithmic or uniform distribution.
The parameters sampled from a logarithmic distribution were $0.025<kT_{\rm disk}$ [keV] $<4$, $10^{-2}<N<10^{10}$, $10^{-3}<f_{\rm sc}<1$, and $0.005<N_H [10^{22} {\rm cm}^{-2}] <75$, while we sampled $1.3<\Gamma<4$ from a uniform distribution.
We generate two synthetic spectra datasets, with and without noise, as the decoder does not need to learn to reproduce noise; however, training the encoder with Poisson noise can help improve the robustness.

\begin{figure}
    \includegraphics[width=\columnwidth]{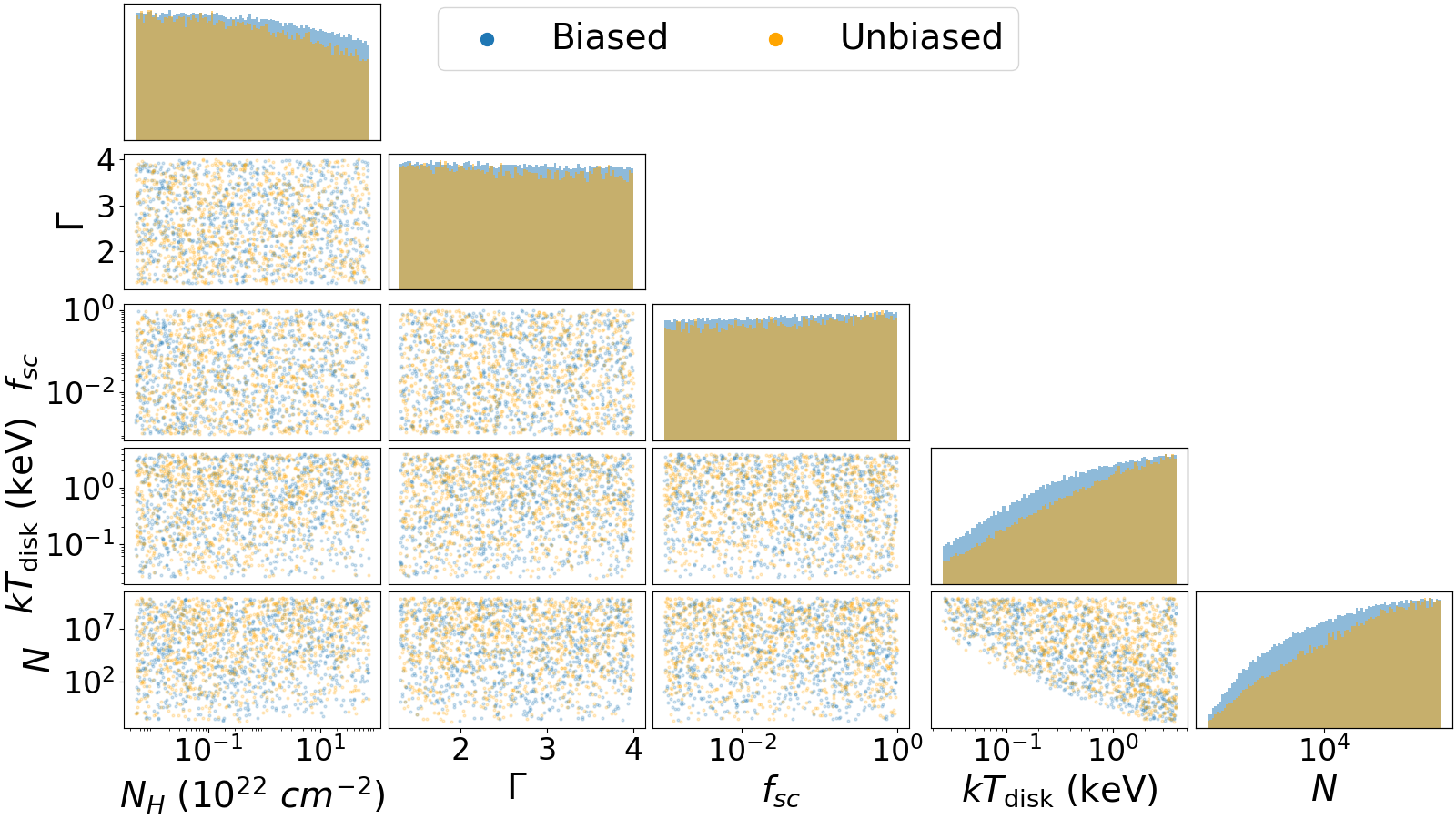}
    \caption{
        Distribution (diagonal histograms) of biased (blue) and unbiased (orange) synthetic parameters and correlations (scatter plots) for each free parameter.
        Each distribution would ideally be flat; however, constraints on the generated spectra cause the parameter distributions to be non-uniform.
        The biased data increases the probability of undersampled data to counter the restrictions on allowed parameters.
    }
    \label{fig:synth_pair_plot}
\end{figure}

After generating each spectrum, we checked if the sum of the counts was above a threshold value of $10^4$ and that there were no integer overflows due to the maximum count exceeding the maximum value of a signed 32-bit integer.
However, this minimum threshold leads to a bias towards parameters that produce higher counts.
We weighted the parameter probability distribution to favour undersampled values to minimise this bias.
We show in Figure~\ref{fig:synth_pair_plot} the distribution (diagonal histograms) and correlation (scatter plots) of synthetic parameters without anti-bias weighting (orange) and with anti-bias weighting (blue).
The Figure shows that the anti-bias weighting does help increase the undersampled values, particularly in parameters $kT_{\rm disk}$, which sees an increase in counts up to 50\%, or 42\% for the smallest values, and $N$, which sees an increase up to 80\%, or 74\% for the smallest values.
$N_{H}$ also benefits from the anti-bias, with the highest values showing an increase in counts by 9.8\%.
However, the correlation plot between $N$ and $kT_{\rm disk}$ shows that part of the parameter space is inaccessible for our imposed constraints.
We repeated this process until we generated 100,000 spectra and then applied the same preprocessing techniques mentioned in section~\ref{sec:preprocessing}.

\subsection{Network Architecture}

\begin{figure*}
    \includegraphics[width=\textwidth]{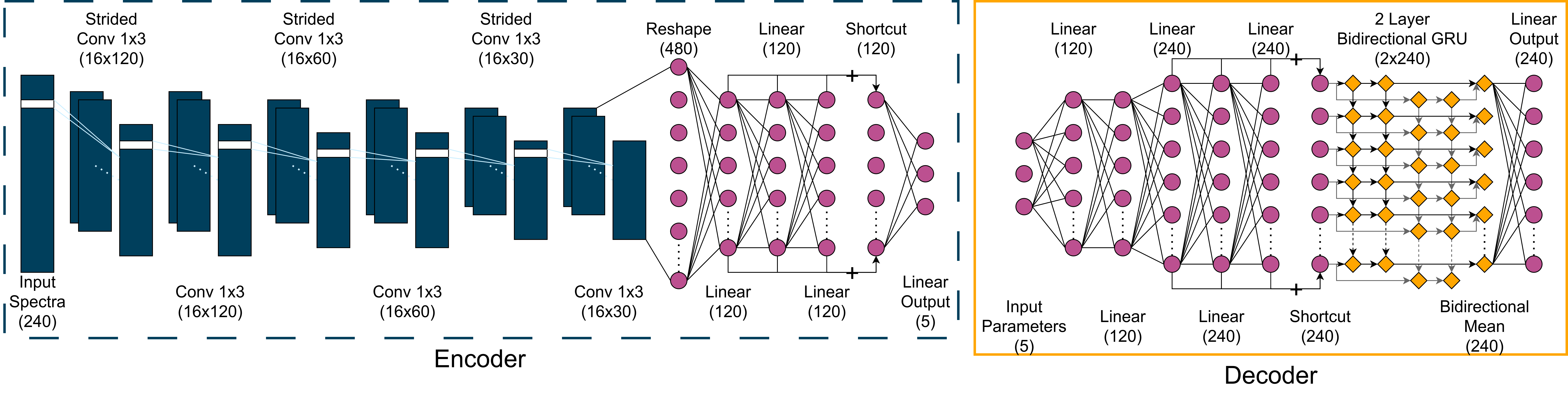}
    \caption{
        Our autoencoder architecture with the encoder (blue dashed box, left) and decoder (orange box, right).
        The text describes each layer, with the column representing the layer.
        The layers are linear (pink circles), convolutional (blue rectangles) and recurrent (orange diamonds).
        The recurrent layer is a 2-layer bidirectional gated recurrent unit with black arrows representing the forward direction and grey arrows representing the reverse.
        Shortcut connections are shown by black arrows with a plus sign.
    }
    \label{fig:autoencoder}
\end{figure*}

Figure~\ref{fig:autoencoder} shows the architecture developed for this project.
The encoder (blue dashed box, left) learns the spectra mapping to spectral parameters, while the decoder (orange box, right) learns the inverse operation.
The text describes the layers in the network, with the column representing the layer.
We show linear layers as pink circles, convolutional as blue rectangles and recurrent as orange diamonds.
Black arrows with a plus sign represent shortcut connections where we add outputs from the connected layers \citep{he2016deep}.
The convolutional layers alternate using strided convolution, where the stride is two to downscale the input length by a factor of two, and regular convolution, where the input length is preserved \citep{springenberg2014striving}.
The recurrent layer is a two-layer bidirectional gated recurrent unit \citep{schuster1997bidirectional} with arrows depicting the flow direction; the recurrent's output is the average of the two directions.

We use scaled exponential linear unit activation in each linear layer and exponential linear units for convolutional and recurrent layers to introduce non-linearity.
These activations help to prevent gradient instabilities during backpropagation by pushing the outputs of layers towards a mean of zero for exponential linear units \citep{clevert2015fast} or a normal distribution for scaled exponential linear units \citep{klambauer2017self}.

The first five layers of the decoder are linear because they enable non-linear mappings of parameter space to spectral space and expand the low-dimensional input to the high-dimensional output.
The first two linear layers have 120 neurons, while the remaining three have 240 neurons each to balance network accuracy and computational time, as layers closest to the input have less information to predict, while deeper layers can learn more complex patterns.
The shortcut connection across the three 240-neuron layers improves gradient flow during backpropagation and allows the network to minimise the importance of layers if the extra complexity is unnecessary.
The linear layer after the recurrent layer improves the network's accuracy by allowing the linear layer to weight the output from the recurrent features.

We use a two-layer bidirectional recurrent layer to learn the sequential relationships of spectra as the bi-directionality helps the network link the behaviour between high and low energies instead of only providing uni-directional constraints.
We use a gated recurrent unit over traditional recurrent layers, which suffer from vanishing gradients, and long short-term memory \citep{hochreiter1997long}, which are more computationally expensive and offer negligible performance improvements for small input sizes.

Similar reasoning is applied when designing the encoder; however, we use convolutional rather than recurrent layers to identify spatial relationships of spectra.
Each convolutional layer has a kernel size of three and uses 16 filters, and the strided convolution allows the convolution kernels to have different receptive fields.
We further explain our choice in network designs in section~\ref{sec:development}.

We used mean squared error (MSE) loss functions with the Adam optimiser \citep{kingma2014adam} for stochastic gradient descent with a $5\times10^{-4}$ learning rate and a weight decay of $10^{-5}$; we found these produced the most stable training for the decoder and encoder.
We initially considered using {\sc PyXspec} fit statistics as the loss function to judge the encoder's performance; however, this approach proved inefficient due to the time costs of loading spectra with response files into {\sc PyXspec} and the inability to accelerate the performance using graphics processing units (GPUs).

To further improve training convergence, we implemented a scheduler that reduces the learning rate when the training loss plateaus.
Specifically, if the validation loss did not decrease by more than 10\% within ten epochs, the learning rate would decrease by a factor of two.
This scheduler allows a higher initial learning rate to converge quickly on the optimal solution, avoiding local minima and then reducing the learning rate to converge on the minimum.

\subsection{Training}

We split the supervised and unsupervised datasets into training and validation sets in a 9:1 ratio to prevent the network from overfitting the data.
During training, we updated the network's parameters using backpropagation on the training data while assessing its performance on the validation data and tuning hyperparameters to improve its generalisation to new data.

We trained both halves of the autoencoder for 200 epochs.
Then, we evaluated the accuracy of the trained encoder on the validation dataset using the reduced ``PGStat'' in {\sc PyXspec} (a fit statistic for Poisson-distributed data with Gaussian-distributed background).
We are using the 3C50 background model\footnote{\url{https://heasarc.gsfc.nasa.gov/docs/nicer/analysis_threads/scorpeon-overview/}}, which has systematic uncertainties that we assume are Gaussian, on top of this; as seen in section~\ref{sec:synth_spectra}, for training the decoder, we randomise the background, further introducing stochasticity; therefore, goodness of fit tests have limited utility due to the shortcomings and imperfect calibrations of the background model.
Assuming a perfect model and data, the reduced PGStat should be $\sim 1$; however, we do not expect to achieve this ideal value due to our choice of a simplistic model.

We compared the encoder's predictions with five different configurations: (i) the encoder's predictions alone, (ii) the encoder's predictions with ten iterations of {\sc PyXspec} curve fitting, (iii) previously optimised fit-parameters derived from brute-force {\sc PyXspec} spectral fits, (iv) fiducial parameter settings, and (v) ten iterations of curve fitting using {\sc PyXspec} starting from fiducial parameter settings.
Fiducial parameters are the default parameter values in {\sc PyXspec}; therefore, they are unrelated to the data, so they typically match the data very poorly; accordingly, this evaluation sets a benchmark ceiling reduced PGStat as a reference point.

\subsection{Network Interpretability}

Neural networks are infamous for being black box functions; therefore, we utilise layer visualisation and saliency maps to understand how the network makes its predictions.

Layer visualisation helps us identify what the network is learning and what features each layer finds meaningful.
During the network's architecture design, we visualised the output from each layer to identify the importance and how different layer types interact.
For the fully connected layers closest to the latent space, the hidden outputs will not have much physical resemblance; however, we can view the mappings of hidden inputs to hidden outputs through the weights used.
We can directly plot the outputs from the convolutional and recurrent layers, which should resemble higher-level spectral features.

We used saliency maps \citep{simonyan2013deep} to identify which inputs the network focuses on most to improve itself.
Saliency maps are the gradients of the inputs relative to the loss function and are calculated using backpropagation, the same method to update the network's weights, where the magnitude of the gradients indicates the sensitivity to a particular input.
We calculated the saliency maps for both the decoder, where the inputs are parameters, and the autoencoder, where the inputs are spectra.


\section{Results}

\subsection{Model Development}
\label{sec:development}

We investigate several architectural designs and present the four main decoder architectures: (i) a six-layer linear network, (ii) a nine-layer convolutional network, (iii) a seven-layer recurrent network, (iv) and a ten-layer recurrent-convolutional network.
After finding the optimal architecture, we perform further training and optimisation to obtain the best performance for the decoder.

\begin{figure*}
    \includegraphics[width=0.9\textwidth]{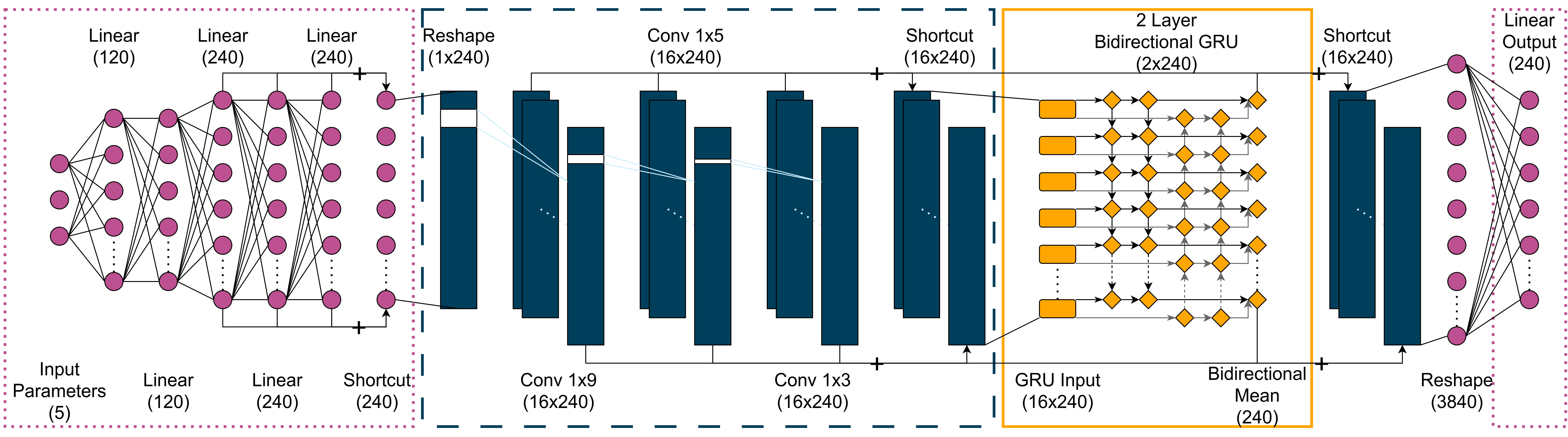}
    \caption{
        The decoder architectures for the linear network (pink dotted boxes), convolutional network (pink dotted and blue dashed boxes), recurrent network (pink dotted and orange boxes) and recurrent convolutional network (all layers).
        The text describes each layer, with the column representing the layer.
        The layers are linear (pink circles), convolutional (blue rectangles) and recurrent (orange diamonds).
        The recurrent layer is a 2-layer bidirectional gated recurrent unit with black arrows representing the forward direction and grey arrows representing the reverse.
        Shortcut connections are shown by black arrows with a plus sign.
    }
    \label{fig:decoder_architecture}
\end{figure*}

Figure~\ref{fig:decoder_architecture} shows the architectures.
Following the design style in Figure~\ref{fig:autoencoder}, linear layers are pink circles, convolutional are blue rectangles, recurrent are orange diamonds, and black arrows with a plus sign represent shortcut connections.
All layers within the pink dotted boxes comprise the linear network.
In addition to the pink dotted box, layers within the blue dashed box comprise the convolutional network, and layers within the orange box comprise the recurrent network.
The overall architecture shown comprises the recurrent convolutional network.
The linear and recurrent networks use a dropout \citep{srivastava2014dropout} probability of 1\%, and convolutional and recurrent convolutional networks use a dropout probability of 2\%.

In constructing the linear network, we found that using several layers of neurons equal to the number of spectral bins worked better than gradually increasing the number of neurons up to the spectral dimensionality; however, more neurons increase the chance of overfitting.
We also observed that increasing the number of layers improved accuracy and training stability, but accuracy gains plateaued, and processing time and overfitting increased.
We found that two 120 and four 240-neuron linear layers provided the most accurate results on the validation dataset, balancing training accuracy and overfitting.

To reduce overfitting, we added dropout with a 1\% probability on all layers except the first and last layers.
Because those layers are critical for initially inferring information from the input or reconstructing the spectrum, they are particularly sensitive to dropout; therefore, we opted against its use.
Setting the dropout probability too high caused the network to struggle to learn and plateau too early, leading to poor validation accuracy and instability.
Setting the probability too low led to overfitting and high training accuracy but poor validation accuracy.

To construct the convolutional network, we utilised three convolutional layers with varying kernel sizes of 9, 5, and 3 to capture spatial features on different scales, with each layer having 16 filters to balance accuracy and computational time.
These convolutional layers were placed before the final linear layer to improve the network's accuracy by allowing the linear layer to weight the output from the convolutional layers.
Batch normalisation \citep{ioffe2015batch} was applied only to the first convolutional layer to renormalise the hidden outputs and prevent vanishing or exploding gradients; however, we found this introduces training instabilities.

The shortcut connection between the convolutional and recurrent layers in the recurrent convolutional network allows the network to balance the two methods.

Table~\ref{tab:network_comparison} shows the average MSE of five runs trained for 100 epochs on real data and the training time for one epoch for each decoder.
Our best-performing network was the recurrent network, with the highest accuracy, joint second-best training time, and most stable training.
The linear network has the fastest training time but the lowest accuracy, the most significant variability in accuracy and often suffers from overfitting.
Convolutional and recurrent convolutional networks obtain similar accuracies; however, recurrent convolution is the slowest, while the convolutional network has the most consistent final loss but the most significant fluctuations in loss per epoch.
The loss functions for one training run per network can be seen in Appendix~\ref{appendix:decoder_loss}.

\begin{table}
    \centering
    \caption{
        Comparison between the time to train for a single epoch and the average MSE of five runs after training for 100 epochs for the different decoder architectures.
        The decoder architectures are a six-layer linear network, a nine-layer convolutional network, a seven-layer recurrent network and a ten-layer recurrent convolutional network.
    }
    \label{tab:network_comparison}
    \begin{tabular}{*{3}{c}}
        \hline
        Network Name & Training Time & MSE\\
        --- & $s\ {\rm epoch}^{-1}$ & $10^{-4}$\\
        \hline
        Linear & \textbf{0.4} & $7.19 \pm 1.30$\\
        \hline
        Convolutional & 0.7 & $6.44 \pm 0.74$\\
        \hline
        Recurrent & 0.7 & $\boldsymbol{5.45} \pm 1.02$\\
	\hline
	Recurrent convolutional & 1.0 & $6.74 \pm 1.30$\\
        \hline
    \end{tabular}
\end{table}

We adopt the recurrent network trained for 200 epochs on synthetic data as our benchmark approach for the rest of the report.
However, as the synthetic data has greater diversity and no noise, the network can generalise better; therefore, we reduce the dropout probability to 0.1\% and remove dropout from one of the linear layers.
The recurrent network has a final MSE of $1.06 \times 10^{-5}$.

Next, we investigate six scenarios in which we compare the training method for the encoder of supervised learning vs semi-supervised.
We also compare the network's performance on synthetic and real spectra with fitted parameters.
The three training scenarios are:

\begin{itemize}
	\item Encoder trained with an MSE loss on the parameters for a supervised approach.
	\item Encoder trained using the reconstructions loss from a previously trained decoder for an unsupervised approach.
	\item Combination of the previous two where the parameter loss has a relative weighting of $0.01$ for a semi-supervised approach.
\end{itemize}

In the unsupervised and semi-supervised approach, the previously trained decoder is used, which helps enforce the physical latent space; therefore, the overall autoencoder is still trained using semi-supervised learning.
We test each training scenario with the encoder trained on synthetic and real spectra and evaluate the performance by testing the fits using {\sc PyXspec} to get the reduced PGStat.
Due to variations in the real spectra, we fixed the training and validation datasets for all scenarios.
These variations can drastically change the reduced PGStat, which does not reflect the network's accuracy.
The variations between the datasets and the learning methods can result in the encoder overfitting or underfitting too much; therefore, for real spectra, we use a dropout of 10\% and a learning rate of $5\times 10^{-4}$, and for synthetic spectra, we use a dropout of 1\% and a learning rate of $10^{-4}$.
We apply dropout to all convolutional and recurrent layers; however, we only use dropout on one linear layer for real supervised and semi-supervised learning and none for the rest.

Table~\ref{tab:training_comparison} shows the median reduced PGStat of five runs trained for 100 epochs.
The best performance for training the encoder is consistently achieved using real spectra, with an improvement factor of 1.2-5.3 over results obtained with synthetic spectra.
For synthetic spectra, unsupervised learning performs the best, whereas supervised learning performs significantly worse by a factor of 6.7.
When using real spectra, the three approaches are more consistent, with semi-supervised learning achieving the best reduced PGStat and supervised learning performing the worst by a factor of 1.9.

\begin{table}
    \centering
    \caption{
        Comparison of the median reduced PGStat of five runs after training for 100 epochs for different training methods of the encoder.
        Two factors are compared: the loss function, which can be supervised, unsupervised or semi-supervised, and the training dataset for the encoder, which can be synthetic or real.
    }
    \label{tab:training_comparison}
    \begin{tabular}{*{4}{c}}
        \hline
        & supervised & unsupervised & semi-supervised\\
        \hline
        Synthetic Spectra & $1190 \pm 380$ & $178 \pm 49$ & $362 \pm 64$\\
        \hline
        Real Spectra & $225 \pm 25$ & $148 \pm 59$ & $\boldsymbol{117} \pm 26$\\
        \hline
    \end{tabular}
\end{table}

For the rest of the report, we will use the encoder trained for 200 epochs on real data using semi-supervised learning.
The final MSE for the reconstruction and latent space is $1.45\times 10^{-3}$, and the reduced PGStat is 62.7.

\subsection{Model Evaluation and Analysis}

We evaluated the saliency of the decoder and autoencoder on the validation dataset.
The decoder's saliency produces five values per spectrum, indicating the gradient for each parameter of the spectral model with respect to the loss function.
We take the average ($S_\mu$) and standard deviation ($S_\sigma$) over the dataset and normalise them by the smallest $S_\mu$, as the relative weights are what is important, shown in table~\ref{tab:decoder_saliency}.
$\Gamma$ and $f_{\rm sc}$ show the smallest saliencies; therefore, these parameters have the lowest impact on the network's prediction.
In contrast, $kT_{\rm disk}$ and $N$ show the greatest saliencies, influencing the network's prediction the most; however, they also show the most significant variance.

\begin{table}
    \centering
    \caption{Saliency calculated from the differentiation of the decoder's loss function with respect to each free spectral parameter averaged over all validation spectra ($S_\mu$) and the standard deviation of the saliencies ($S_\sigma$) normalised by the smallest $S_\mu$, as the absolute value is arbitrary.}
    \label{tab:decoder_saliency}
    \begin{tabular}{*{6}{c}}
        \hline
        & $N_H$ & $\Gamma$ & $f_{\rm sc}$ & $kT_{\rm disk}$ & $N$ \\
        \hline
        $S_\mu$ & 2.97 & 1.00 & 1.22 & 8.92 & 8.83 \\
        \hline
        $S_\sigma$ & 5.93 & 1.70 & 1.72 & 9.26 & 8.06 \\
        \hline
    \end{tabular}
\end{table}

\begin{figure}
    \includegraphics[width=\columnwidth]{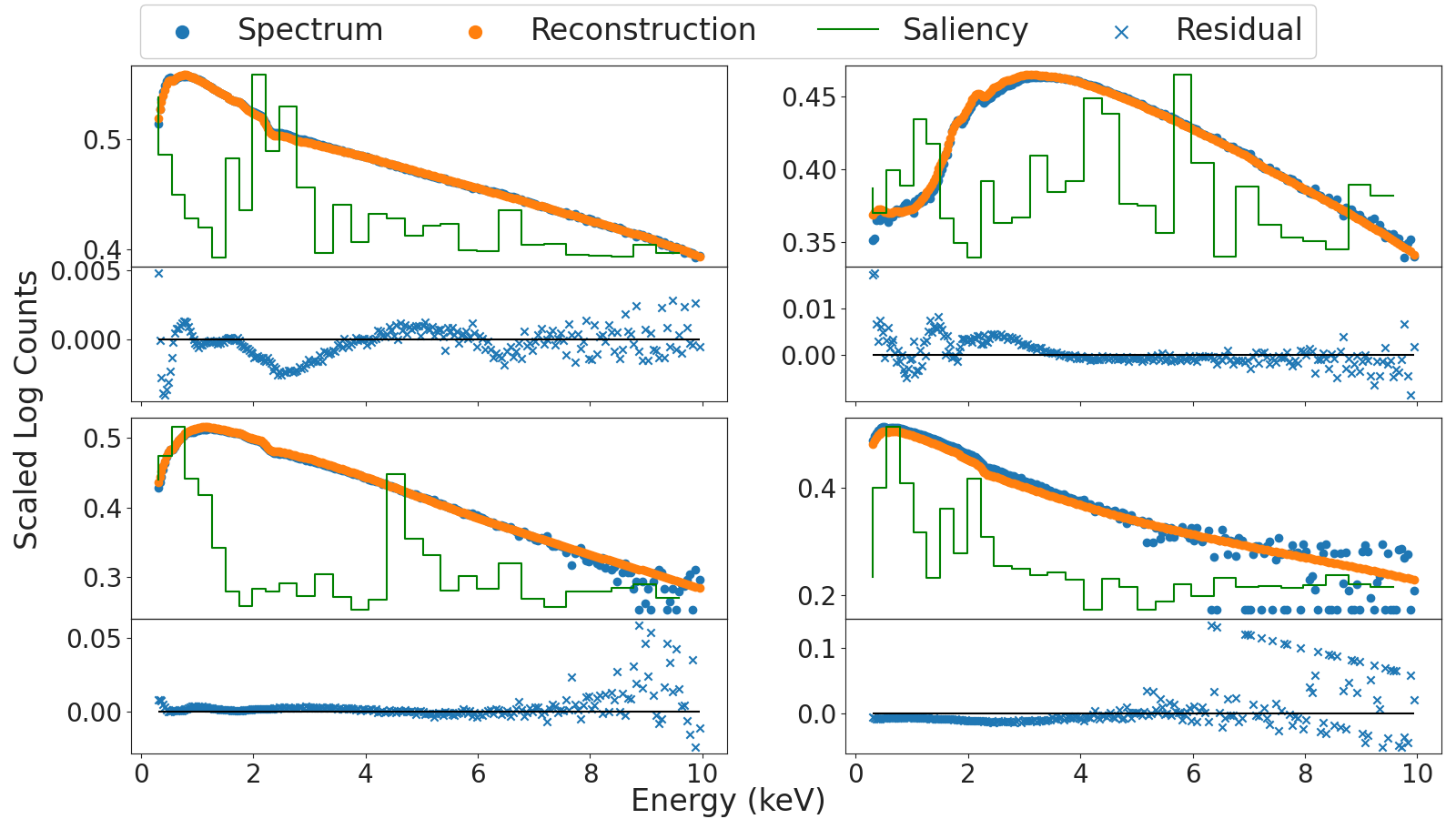}
    \caption{
        A plot of scaled logarithmic count rates against energy (top subplot) for the spectrum (blue), reconstruction (orange), saliency (green line, arbitrary y-scale), and the residuals (bottom subplot) for four different examples.
        The top right plot shows a harder spectrum where the peak is shifted to higher energies, while the other three are softer spectra where the MTD is the dominant component.
        The horizontal line artefacts seen in the spectra of the bottom two plots are due to the logarithmic transform of zero count values, so they are assigned the lowest positive value, resulting in the horizontal lines.
    }
    \label{fig:autoencoder_saliency}
\end{figure}

The autoencoder's saliency has the same dimension as the spectra, so we show four examples.
We binned the encoder's saliency into 30 regions to more easily visualise the overall trend, as shown in Figure~\ref{fig:autoencoder_saliency}.
The Figure shows the reconstructed spectrum (orange) and saliency (green line) alongside the original spectrum (blue) and residuals (blue) compared to a zero line (black).
The x-axis represents the energy in keV, the y-axis represents the scaled logarithmic counts for the spectra and residuals, and the y-axis is arbitrary for the saliency.
The horizontal line artefacts seen in the spectra of the bottom two plots are due to the logarithmic transform of zero count values, so they are assigned the lowest positive value, resulting in the horizontal lines.

The autoencoder saliencies generally place the most importance on the MTD peak around 0-2 keV.
For the softer spectra (left column and bottom right), the network also focused on the point where the power law began to dominate the MTD flux, around 2 keV (top left and bottom right), or at higher energies in the 4-7 keV range (bottom left).
Generally above 7 keV, the network places less significance on the values.
For the harder spectrum (top right), the saliency is more evenly distributed throughout the energy range.

\begin{figure}
    \includegraphics[width=\columnwidth]{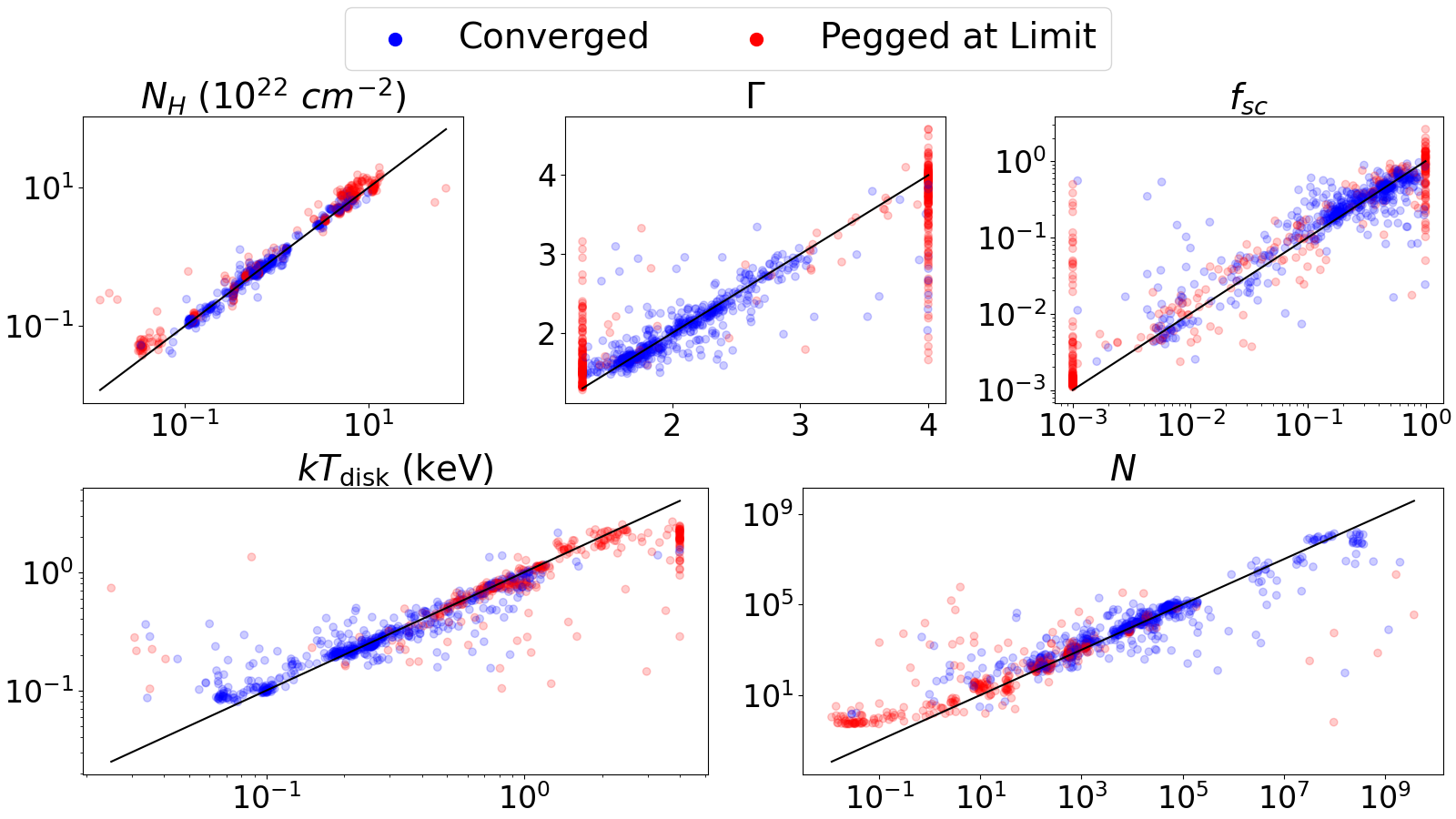}
    \caption{
        Comparison between the encoder's parameter predictions (y-axis) and precalculated parameters (x-axis) in logarithmic space, with a black line to show perfect agreement.
        Spectra with at least one precalculated parameter pegged at a minimum or maximum value during traditional fitting are generally less well-constrained and are coloured red; otherwise, they are blue.
        Pegged parameters can be seen by artificial vertical lines most commonly found in $\Gamma$, $f_{\rm sc}$.
    }
    \label{fig:parameter_comparison}
\end{figure}

Next, we compared the parameter predictions obtained from the encoder with those obtained using traditional methods with {\sc PyXspec}, as shown in Figure~\ref{fig:parameter_comparison}.
The black line shows perfect agreement between the two methods, and the red values are where at least one parameter for a given spectrum is pegged at a minimum or maximum value, as these are generally less well-constrained.

We calculate $\chi_\nu^2$ between the parameters with uncertainties obtained from the traditional fitting to compare the agreement between the five parameters.
The $\chi_\nu^2$ values are 0.721 for $N_H$, 0.909 for $\Gamma$, 1.71 for $f_{\rm sc}$, 29.4 for $kT_{\rm disk}$, and 21.9 for $N$.
We use bootstrapping to calculate the lower limit of the $4\sigma$ $\chi_{\nu}^2$ null hypothesis leading to the values 124 for $N_H$, 38.4 for $\Gamma$, 48.4 for $f_{\rm sc}$, 205 for $kT_{\rm disk}$, and 208 for $N$, all of which are significantly larger than the previously calculated $\chi_\nu^2$; therefore, the null hypothesis can be rejected.
The comparisons show that the MTD parameters were the least well-modelled, while $\Gamma$, $f_{\rm sc}$, and $N_H$ have similar values.
The power law parameters and $kT_{\rm disk}$ have significant deviations due to hard limits at the upper and lower boundaries.

\begin{figure*}
    \includegraphics[width=\textwidth]{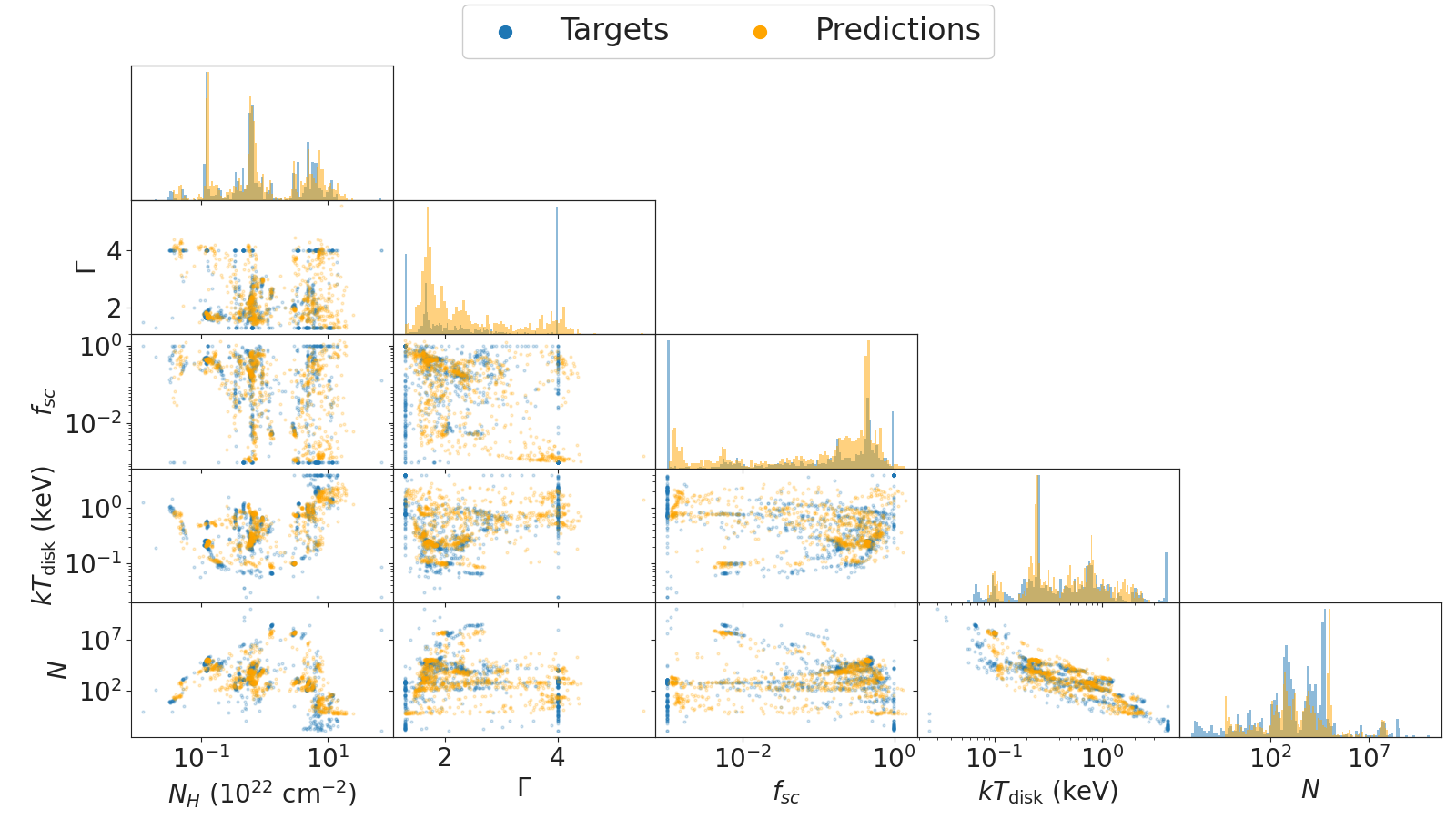}
    \caption{
        Distribution (diagonal histograms) of encoder predictions (orange) and precalculated parameters (blue) and correlations (scatter plots) for each free parameter in the model.
        Boundary conditions are shown by the vertical peaks at the edges of the histogram distributions, particularly seen in $\Gamma$ and $f_{\rm sc}$.
        $kT_{\rm disk}$ and $N$ are the only parameters to show a correlation.
    }
    \label{fig:pair_plot}
\end{figure*}

Figure~\ref{fig:pair_plot} shows the distribution (diagonal histograms) and correlation (scatter plots) of precalculated parameters (blue) and encoder predictions (orange).
The encoder correctly predicts the parameter distributions, except for $kT_{\rm disk}$ and $N$, which struggle to capture the distribution range at either extremum.
The correlation plots reveal no significant correlations between the parameters, except for a strong negative correlation in logarithmic space between $kT_{\rm disk}$ and $N$.
The upper and lower limits of $\Gamma$ and $f_{\rm sc}$ and the upper limit of $kT_{\rm disk}$ produce spikes in the precalculated parameter distribution due to the hard limits imposed by the model.
However, the encoder's predictions display tails instead of reflecting these hard limits.

The final part of this section will compare different fitting methods using {\sc PyXspec} to calculate the computational time and reduced PGStat for five scenarios: the encoder's predictions, the precalculated parameters\footnote{The fitting was performed on 5-10 cores of an AMD 6376 \citep{amd20136376}; therefore, the precise computational time is not comparable.\label{alternate_hardware}}, fiducial parameters for the model, the encoder's predictions with ten iterations of {\sc PyXspec} fitting, and fiducial parameters with ten iterations of {\sc PyXspec} fitting.
Fitting using {\sc PyXspec} is multithreaded using eight threads from an Intel 
i7-10750H central processing unit (CPU) \citep{intel2020i7}.
The encoder's parameter prediction uses an NVIDIA mobile 1650 ti GPU \citep{nvidia20201650}.
Table~\ref{tab:pyxspec_results} shows the averaged time from five runs to generate the parameter predictions and the median reduced PGStat for the 1,080 spectra in the validation dataset for each fitting method.

\begin{table}
    \centering
    \caption{
        Comparison of computational time and reduced PGStat for different fitting methods.
        Scenarios are predictions from the encoder, precalculated parameters as the goal to reach, fiducials as the benchmark ceiling, an encoder with ten iterations of {\sc PyXspec} fitting, and ten iterations of {\sc PyXspec} from the fiducial values.
        The encoder's parameter predictions use an NVIDIA GPU.
    }
    \label{tab:pyxspec_results}
    \begin{tabular}{*{3}{c}}
        \hline
        Scenario & Time & Reduced PGStat\\
        --- & s & ---\\
        \hline
        Encoder & $(\boldsymbol{3.61} \pm 0.41)\times 10^{-2}$ & 62.7\\
        \hline
        Precalculated\footref{alternate_hardware} & $\sim 1.8\times 10^4$ & \textbf{4.44}\\
        \hline
        Fiducials & 0 & $1.15\times 10^4$\\
        \hline
        Encoder + Fitting & $96.2 \pm 2.0$ & 4.76\\
        \hline
        Fiducials + Fitting & $97.7 \pm 0.9$ & 77.4\\
        \hline
    \end{tabular}
\end{table}

All optimisation methods achieve reduced PGStat values much lower than the fiducial threshold.
The precalculated parameters have the lowest reduced PGStat but take the most computational time, having been fitted by brute force.
On the other hand, the encoder's predictions have a significantly lower computational time but at the cost of a higher reduced PGStat.
The encoder achieves an accuracy similar to ten fitting iterations; however, it improves the computational time by a factor of 2,700.
If, instead, we run the encoder on the CPU, the computational time becomes $0.273 \pm 0.008$ s, 7.6 times slower than the GPU but 360 times faster than {\sc PyXspec} for similar accuracy.
However, using the encoder with ten fitting iterations brings the reduced PGStat close to the precalculated values.
To achieve a similar reduced PGStat as the encoder with ten {\sc PyXspec} fitting iterations requires 60 fitting iterations from the fiducial values and takes 288 seconds, resulting in a three-fold improvement in computational time.


\section{Discussion}

\subsection{Model Development}
\label{sec:discuss_develop}

Table~\ref{tab:network_comparison} shows that all decoders achieve similar accuracies, which could be due to their similar architectures, leading to the conclusion that linear layers are crucial in effectively mapping spectral parameters into corresponding spectra.
Convolutional and recurrent layers provide accuracy improvements ranging from 10-30\%; however, it is necessary to consider other factors such as training time, stability, and consistency when choosing an architecture.

The recurrent convolutional network architecture combines the recurrent and convolutional layers, which leads to improved accuracy over the linear network.
However, this approach performs worse than the recurrent or convolutional networks while being more computationally expensive.
The recurrent convolutional network also has the joint highest variance in the final loss due to it sometimes overfitting, resulting in a significantly higher loss than on other runs.
The degradation in performance may be due to the different layer types disrupting the features that other layers are learning, thereby breaking the spatial relationships learned by the convolutional layers and the sequential relationships learned by the recurrent layers.

The convolutional network provides a 12\% improvement over the linear network and has the most consistent final loss of the four networks, but it also exhibits the most unstable training.
We can attribute this instability to the batch normalisation process that increases the stochasticity during training.
Meanwhile, the recurrent network architecture achieves the best performance, with a 32\% improvement over the linear network, while having a competitive computational cost and stable training.
Thus, we choose the recurrent network as our adopted optimal architecture.

In the next test, we evaluated different training methods.
We observed that training on synthetic data resulted in the worst performance, while adding the unsupervised loss improved the encoder's accuracy.
The poor performance of training on synthetic data, even though the dataset is larger and more uniform, could be due to the synthetic data being produced using a model that approximates the spectrum of BHBs, which is less accurate when the BHBs are in harder states with more pronounced spectral reflection components, which consists of features and interplay not accurately captured by our model, and accordingly, making the network less robust to real data.
This poor robustness is more evident when we look at the reduced PGStat for supervised learning on synthetic data; the encoder does well when predicting synthetic parameters but is markedly worse on real data.
On the other hand, unsupervised learning did not perform as poorly, likely because of the reconstruction requirement, which is a more challenging problem, acting as a regularisation term, forcing the network to generalise more to prevent overfitting.
The overfitting of supervised learning on synthetic data can also explain why semi-supervised learning performs worse than unsupervised for synthetic but not for real spectra.

For real spectra, we observed that semi-supervised learning performed the best, which could be due to the supervised learning reinforcing the structure of the latent space, coupled with the unsupervised learning enforcing that the parameters can still successfully reconstruct the spectrum.
We found that the encoder predicted parameters outside the training data domain during unsupervised learning.
We speculate that the encoder can exploit regions outside the training domain to achieve good reconstructions but at the cost of a loss of correspondence between the latent space and the physical parameters.
In contrast, we observed that supervised learning could reproduce the target parameters well, but the reconstructions would be worse; for example, if $kT_{\rm disk}$ and $N$ were slightly underpredicted, the accuracy could still be high, but the resulting spectrum's scale would be significantly underpredicted.

\subsection{Model Evaluation and Analysis}

The decoder's saliencies indicate that it mainly focuses on the MTD parameters, possibly due to biases in the model that better represent soft states where the MTD dominates
Alternatively, the decoder might find it more important to focus on low energies as they often contain more information (typically, lower energy channels contain the most X-ray counts), or these parameters are more challenging to predict.

Regarding the MTD, the autoencoder saliency shows that for soft spectra, the network focuses on the MTD peak around 0-2 keV, with a greater focus on the initial rise in counts at the lowest energies and the fall when the spectrum transitions from MTD to power-law dominant above the thermal disk's Wien tail, typically around 2 keV.
Therefore, the network is capturing the shape and broadness of the spectrum; however, it places little emphasis on the peak's magnitude, suggesting that this does not provide additional information that cannot be inferred from the surrounding data.
Soft spectra also show additional peaks in saliency along the power-law component.
These peaks are usually in the 4-7 keV range, where the network may find it helpful to know the amplitude here to inform the gradient and, therefore, the photon index; however, due to the almost constant gradient, the network does not need to apply much focus to this region.

For harder spectra, the saliency is more uniform, with peaks roughly every 2 keV, indicating the network emphasises capturing the power-law component's shape.
The Fe-K$\alpha$ emission line at 6.4 keV is an intriguing feature, as the decoder was trained solely on synthetic data that did not contain any of these lines, and the latent space does not contain any explicit information about the presence of these lines; however, we see that the network ignores this feature, suggesting that it does not substantially affect the parameters.

Comparing the predictions against the precalculated parameters in Figure~\ref{fig:parameter_comparison}, we can observe that the predictions closely agree with the precalculated parameters, with the $N_H$ and power law parameters in greater agreement than the MTD parameters.
The large $\chi_\nu^2$ for $N$ is likely due to the network's difficulty in predicting values over several orders of magnitudes.
As $N$ and $kT_{\rm disk}$ are strongly correlated, the network's difficulty with $N$ could explain why $kT_{\rm disk}$ also has a large $\chi_\nu^2$.
The worse performance for the MTD parameters also correlates with the larger decoder saliencies, suggesting that these parameters are harder for the network to learn.
The power law parameters and $kT_{\rm disk}$ show vertical boundary conditions, $f_{\rm sc}$ has a hard boundary at 0 and 1, while the boundaries for $\Gamma$ and $kT_{\rm disk}$ are more artificial, resulting in the twin peaks seen in Figure~\ref{fig:pair_plot}.
These peaks, shown by red data points in Figure~\ref{fig:parameter_comparison}, create artificial degeneracies that could hinder the network's ability to learn the mapping.
Alternatively, these peaks could be artefacts of the {\sc PyXspec} fitting algorithm, causing the fit to get stuck in local minima and locking these parameters to the boundaries instead of finding the optimal value.
As a result, the network may be conflicted between predicting an accurate parameter far from the precalculated value or minimising the distance to the precalculated value at the cost of the reconstruction, hence the spread in predictions at these boundaries.
The discrepancy between the predictions and precalculated parameters correlates with the relative magnitude of the peaks, further indicating an issue with the {\sc PyXspec} fitting algorithm.
If the problem lies within the {\sc PyXspec} fitting algorithm, using an autoencoder could help avoid these local minima, potentially improving the accuracy of the fits, similar to what \cite{parker2022agn} found.

We compared our network with several different fitting methods in the final test.
Our encoder reduces the computational time by a factor of 2,700 compared to {\sc PyXspec}, as it avoids the computational cost of loading the responses and takes advantage of GPU acceleration.
Even when we use a CPU for the encoder, we still see a factor of 360 improvement for similar accuracy, demonstrating the advantage of this different approach.
However, our current encoder's accuracy is not yet comparable to the best fit from the precalculated parameters.
To address this, we combined the two methods to see if our initial parameter predictions could reduce the computational time while maintaining similar performance.
With ten iterations of {\sc PyXspec}, we achieved similar accuracy to the precalculated parameters.
Fitting from fiducial values required 60 iterations to achieve the same accuracy, increasing the computational time by 3-fold, even though they achieved comparable accuracy to the encoder's predictions from only ten iterations.
We would expect that the additional fitting required for fitting from fiducial parameters would be similar to the encoder; therefore, this could be due to {\sc PyXspec} getting stuck in local minima that the encoder could otherwise help avoid, suggesting an alternate use case for the encoder, where it can help {\sc PyXspec} start from a more suitable location in parameter space and reduce the number of iterations required to achieve high accuracy.

\section{Summary and Future Improvements}

We have presented a novel approach to fitting spectra using semi-supervised autoencoders that significantly reduce the computational time by a factor of 2,700 for establishing initial predictions and a 3-fold speedup in achieving the best fit.
The network can rapidly predict spectral parameters for large datasets to provide starting points for further analysis, reducing computational time and the chance for traditional fitting methods to get stuck in local minima.
We also produce a decoder as a byproduct that can rapidly produce synthetic spectra for provided spectral parameters.
Notably, the current network is rather simplistic and still needs the assistance of traditional fitting to reach a reliable accuracy.
For future work, we will test the network on new BHB data collected by NICER since the summer of 2022 to compare the performance of unseen BHB systems.

One of the main challenges we face is the difficulty of scaling, as spectra and parameters can vary over several orders of magnitude, leading to unstable learning of the network.
To address this, we currently take the logarithm of these values, which can reduce accuracy at high orders of magnitude.
A potential solution is to normalise each spectrum by its minimum and maximum in linear space and have two decoder networks predict the spectrum's shape and minimum and maximum values, respectively.
Passing the scale parameters into a linear layer can ensure that the encoder considers scale when predicting parameters.
This approach can keep the spectra in linear space, allowing the shape decoder to focus more on the high-count parts of the spectrum and improve the robustness of the noisy low-count regions.
However, there could still be significant uncertainty in the scale as the scale decoder would still have its predictions in logarithmic space.

Alternatively, we could use the Poisson uncertainty of the spectra to use a weighted mean squared error as the loss function with the data normalised between the global maximum and minimum in linear space.
The weighted MSE could give the network leeway to match the uncertainty of the data points and improve training stability across several orders of magnitude as the uncertainty would scale with the counts.
However, the large scale could still prove too challenging to train with low-count spectra with values too small for the network to predict.

We would also like to explore the network's performance for different spectral models.
We expect a negligible increase in computational time as model complexity increases, assuming we do not need to make the network deeper.
Suppose the network depth needs to increase for more complex models. In that case, we expect any corresponding increase in computational time to be significantly less than what we would see from traditional fitting. Therefore, the expectation is that for more complex spectral models, the speedups compared to brute-force fitting would be even more profound than reported here.

Another area to improve the accuracy would be to train the model to perfectly fit either a single spectrum that the user wants a good fit for or over a dataset that the user wants parameters for, as opposed to training the network once on a general dataset and evaluating performance on the desired spectrum or dataset.
This method would have the advantage of improved accuracy at the cost of computational time and reusability of the network.
However, instead of training the network from the beginning, we can use transfer learning \citep{bozinovski1976influence}, where we train a generalised pre-trained network on the desired dataset.

Further improvements on the network would be to predict uncertainties in the parameters or the probability distribution for each parameter.
A simple method for obtaining the underlying parameter distribution would be to use Hamiltonian Monte-Carlo \citep{duane1987hybrid}, a faster version of Markov Chain Monte-Carlo \citep{metropolis1953equation} method for differentiable functions (including neural networks).
Another simple method would be to train multiple networks and assess the variance from their predictions; however, this would measure the precision of the network and not necessarily the accuracy.
Another approach would be to use variational autoencoders \citep{kingma2013auto}, which predict the mean and variance of the parameters in the latent space; however, an assumption on the form of the probability distribution is required.

An intriguing avenue for exploration would be to have modular networks that could be combined to form the desired model; for example, we could train three networks: one to predict MTD, another for the power law and another for the Fe-K$\alpha$ emission line.
The user would then be able to choose which modules they want parameters for without the need to train every combination of models.

Furthermore, we could apply our network outside BHBs and use similar observations from NICER with neutron star X-ray binaries, or else consider other X-ray telescopes, or port this approach to other regimes of black hole systems, such as active galactic nuclei.
We are presently testing our architecture on James Webb Space Telescope \citep{gardner2006james} transmission spectra of exoplanets to predict the molecular composition of their atmospheres and on simulated data of the Line Emission Mapper X-ray probe concept \citep{kraft2022line} to predict the redshift of galaxies.



\section*{Acknowledgements}

Thanks to Benjamin Ricketts, Carolina Cuesta-Lazaro, Daniela Huppenkothen, Diego Altamirano, Douglas Burke, Javier Via\~{n}a P\'{e}rez, Jeremy Drake, and the anonymous referee for their advice and support.
This project was also conducted with support from the AstroAI initiative at the Center for Astrophysics | Harvard \& Smithsonian.
This research has made use of data and/or software provided by the High Energy Astrophysics Science Archive Research Center (HEASARC), which is a service of the Astrophysics Science Division at NASA/GSFC.
E.T. and J.F.S. acknowledge support from NASA grant no. 80NSSC21K1886.

\section*{Data Availability}

All data was obtained from the NICER archive: \url{https://heasarc.gsfc.nasa.gov/docs/nicer/nicer\_archive.html}.


\bibliographystyle{mnras}
\bibliography{spectral_fitting}


\appendix

\section{Trial Decoder Losses}
\label{appendix:decoder_loss}

\begin{figure}
    \centering
    \includegraphics[width=\columnwidth]{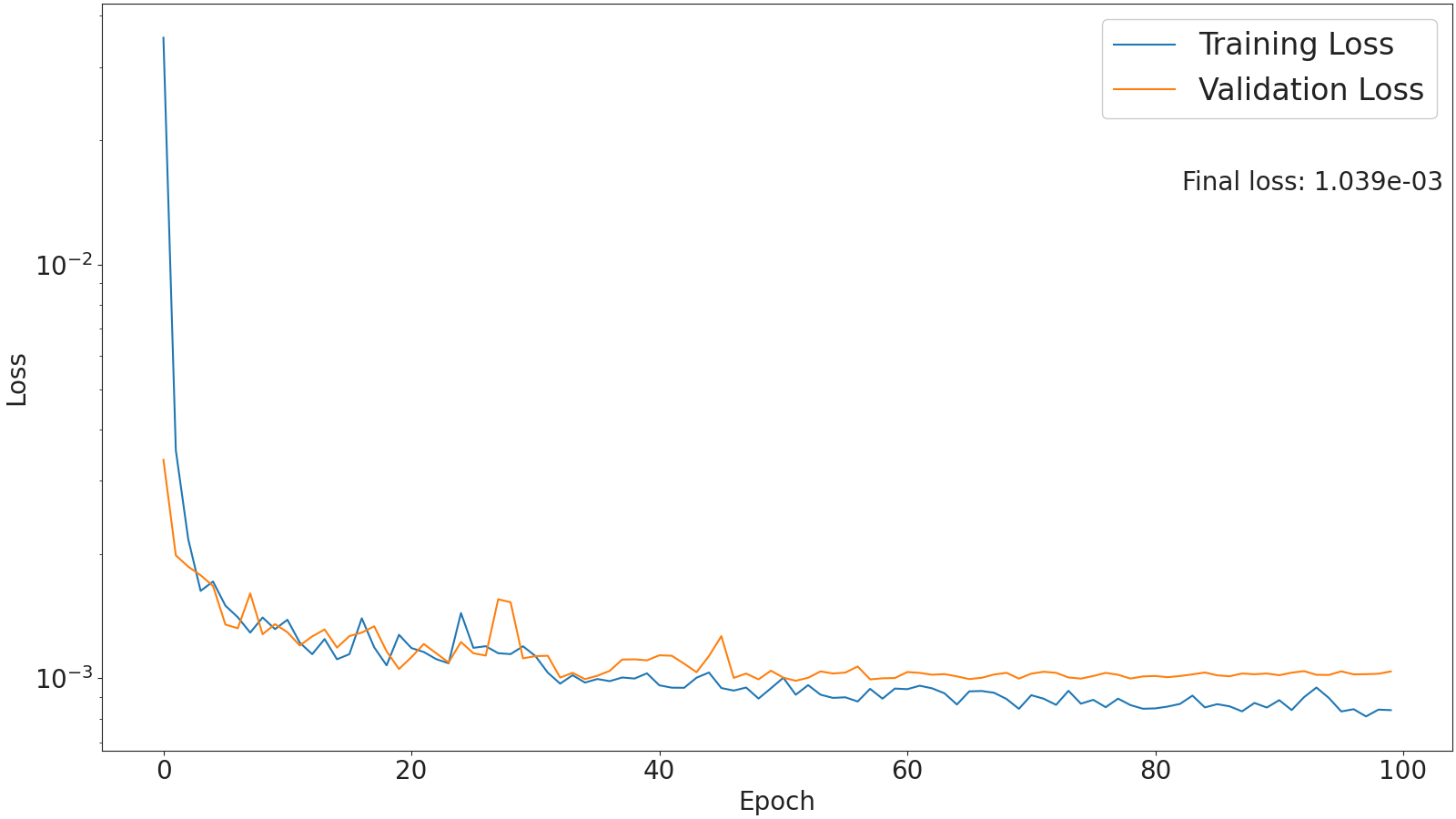}
    \caption{Training (blue) and validation (blue) losses for the linear network during training for 100 epochs.}
    \label{fig:fnn_loss}
\end{figure}

\begin{figure}
    \centering
    \includegraphics[width=\columnwidth]{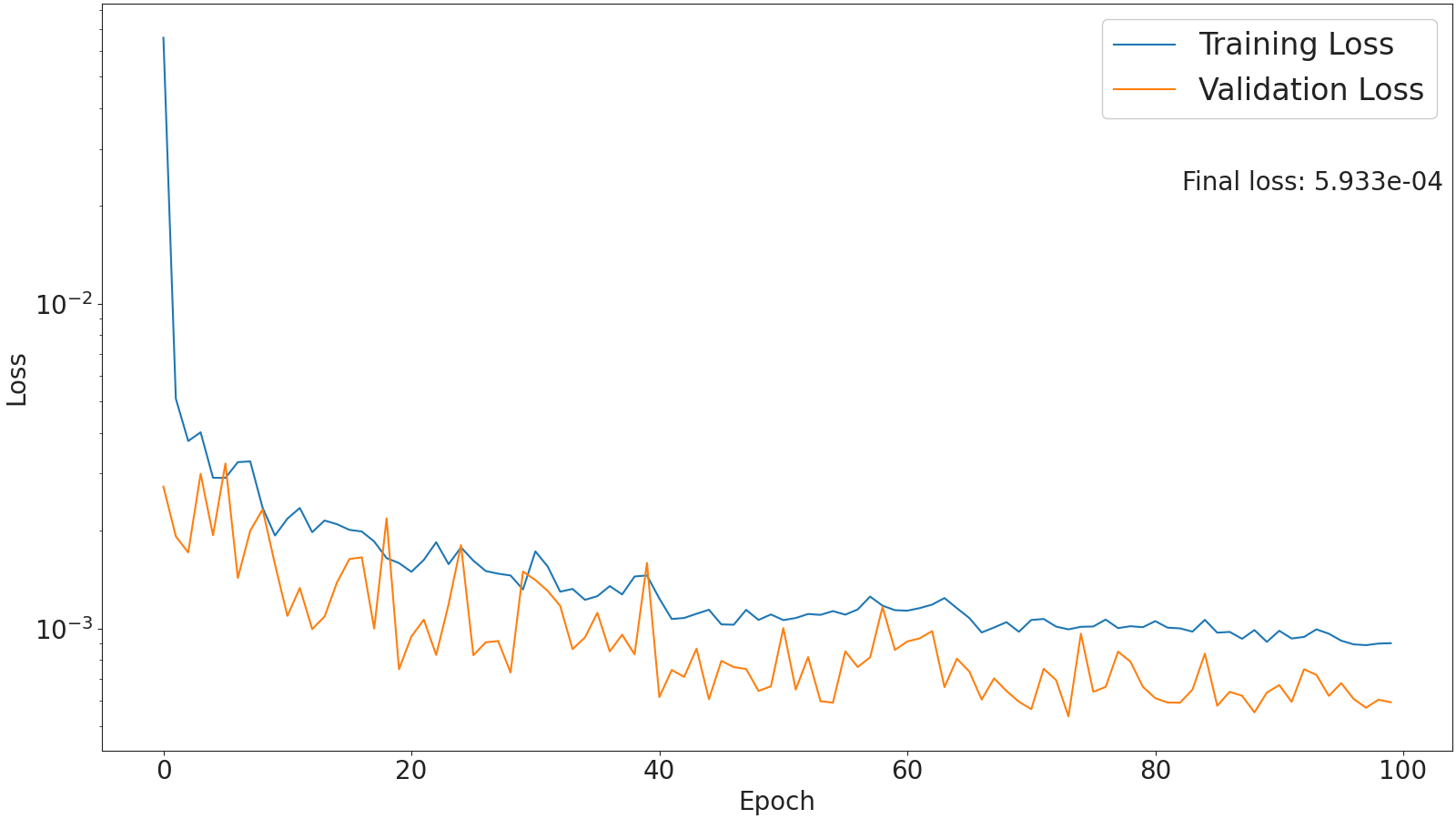}
    \caption{Training (blue) and validation (blue) losses for the convolutional network during training for 100 epochs.}
    \label{fig:cnn_loss}
\end{figure}

\begin{figure}
    \centering
    \includegraphics[width=\columnwidth]{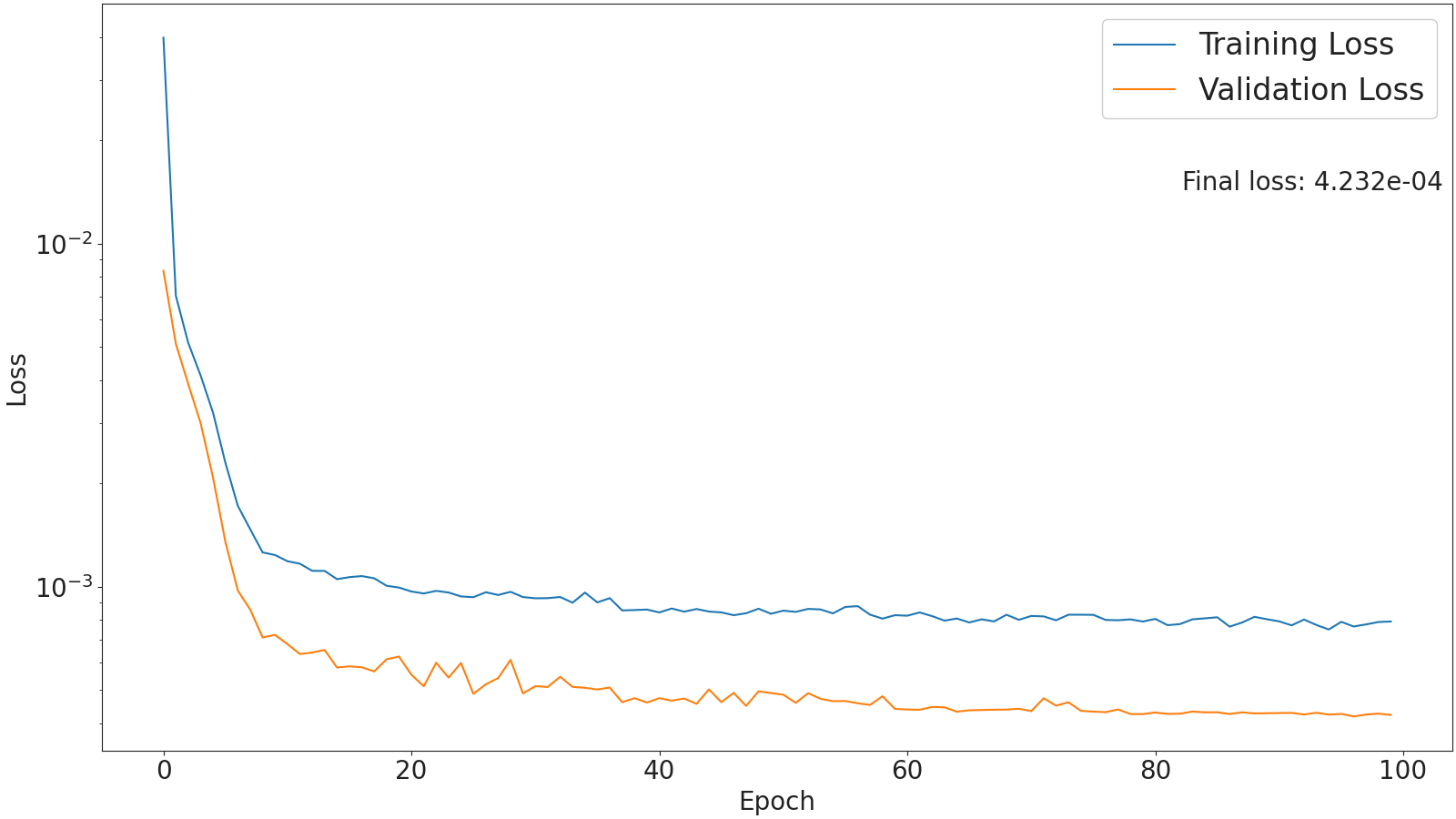}
    \caption{Training (blue) and validation (blue) losses for the recurrent network during training for 100 epochs.}
    \label{fig:rnn_loss}
\end{figure}

\begin{figure}
    \centering
    \includegraphics[width=\columnwidth]{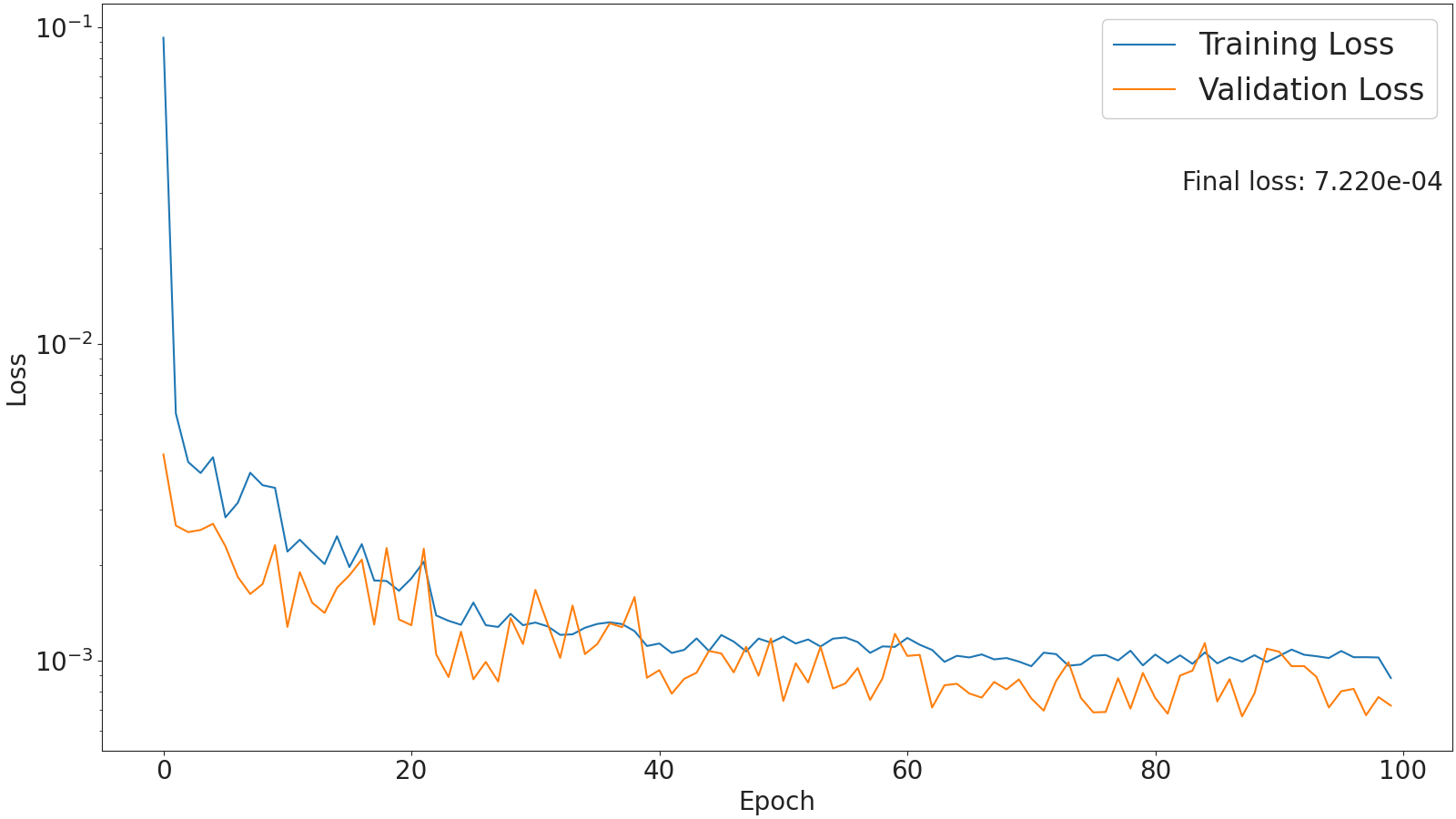}
    \caption{Training (blue) and validation (blue) losses for the recurrent convolutional network during training for 100 epochs.}
    \label{fig:rcnn_loss}
\end{figure}

Figures~\ref{fig:fnn_loss}-\ref{fig:rcnn_loss} show the training (blue) and validation (orange) losses over 100 epochs for the different decoders trained in section~\ref{sec:development}.

Figure~\ref{fig:fnn_loss} shows the training loss for one of the runs of the linear network; however, we can see the network suffered from overfitting, resulting in a significantly higher validation loss of $1.04\times 10^{-3}$ compared to the average result of $7.19\times 10^{-4}$.

Figure~\ref{fig:cnn_loss} shows the training loss for one of the runs of the convolutional network.
We can see that while the network achieved a good final validation loss of $5.93\times 10^{-4}$, the training is unstable, with the validation loss rapidly changing between epochs.
However, the convolutional network can consistently maintain a validation loss less than the training loss, which helps prevent it from overfitting, resulting in a more consistent final loss.

Figure~\ref{fig:rnn_loss} shows the training loss for one of the runs of the recurrent network.
We can see that the training is very stable, and the validation loss is consistently below the training loss, showing that the network is not overfitting the training data.

Figure~\ref{fig:rcnn_loss} shows the training loss for one of the runs of the recurrent convolutional network.
We can see that the network shows similar instabilities as the convolutional network; however, it also has a smaller margin between the training and validation loss, increasing the chance for the network to overfit.

\bsp	
\label{lastpage}
\end{document}